\documentstyle{article}
\newcommand{\vect}[1]{\mbox{\bf #1}}
\newcommand{\vecs}[1]{\mbox{\scriptsize \bf #1}}
\newcommand{\rms}[1]{\mbox{\scriptsize #1}}

\newcommand{\lw}[1]{\smash{\lower 1.5ex\hbox{#1}}}

\begin{document}
% \begin{flushright}
%    hep-ph/9507293
% \end{flushright}
%
%   TITLE & AUTHORS
%
\title{\Large\bf Coulomb corrections for Bose-Einstein correlations
       \vspace{-3mm}\\
       in whole momentum transfer region:\vspace{-3mm}\\
       Proposal of seamless fitting}
\author{M.~Biyajima$^1$\thanks{e-mail: minoru44@jpnyitp.bitnet},
T.~Mizoguchi$^2$, T.Osada$^3$ and G.~Wilk$^4$
\thanks{e-mail: Wilk@fuw.edu.pl}\\
{\small\it $^1$Department of Physics, Faculty of Science,}
\vspace{-4mm}\\
{\small\it Shinshu University, Matsumoto 390, Japan}\vspace{-2mm}\\
{\small\it $^2$Toba National College of Maritime Technology,
Toba 517, Japan}\vspace{-2mm}\\
{\small\it $^3$Department of Physics, Tohoku University, Sendai 980,
Japan}\vspace{-2mm}\\
{\small\it $^4$Soltan Institute for Nuclear Studies, Nuclear Theory
Department}\vspace{-4mm}\\
{\small\it Ho\.za 69, PL-00-681 Warsaw, Poland}
}
\date{\today}
%\date{}
\maketitle
%
%   ABSTRACT
%
\begin{abstract}
We applied an improved Coulomb correction method developed by us
recently to data on identical $KK$-pairs production in S + Pb and $p$
+ Pb reactions at 200 GeV/c obtained by NA44 Collaboration. To
analyse the whole range of the momentum transfers measured the method
of "seamless fitting" has been proposed and used together with the
asymptotic expansion formula for the Coulomb wave function. We found
that such Coulomb corrections lead sometimes to different than
previously reported (by NA44 Collaboration) interaction region and
strongly influence the long range correlations.\\

{\large Preprint {\bf DPSU-95-4} (July, 1995)}
\end{abstract}
\vspace{1cm}
%
%\newpage
{\bf Introduction:}~~~Recently NA44 Collaboration has reported their
data on the Bose-Einstein correlations (BEC) of $K^{\pm}K^{\pm}$ pairs
produced in S + Pb and $p$ + Pb reactions at 200 GeV/c \cite{na4494}.
In our previous work \cite{biya94} we have analysed these data by
making use of the various source functions with the long range
correlation (without, however, invoking any sort of Coulomb
corrections). The data \cite{na4494} have been corrected for Coulomb
interactions by applying only the Gamow factor. As was pointed out
some time ago by Bowler this is, however, not sufficient
\cite{bowler91}. In our recent works \cite{biya94d,biya95} the
still improved method of Coulomb corrections was presented but not
yet applied to any concrete data set. In the present Letter, we apply
it therefore to the analysis of data for $K^{\pm}K^{\pm}$ pairs
production in S + Pb and $p$ + Pb reactions mentioned above (in the
whole measured momentum transfer region) and compare our results with
those obtained before in \cite{biya94}. To be able to analyse the
whole range of measured momentum transfer $Q$ and to avoid wild
oscillations developing at large $Q$'s (cf. Fig. 1a) we have to use
the the asymptotic expansion of the Coulomb wave function together
with the procedure of "seamless fitting" (SF) explained below (cf.
Figs. 1b and 2).\\

{\bf Theoretical formula of BEC with Coulomb wave function:}~~~To
write down an amplitude $A_{12}$ satisfying Bose-Einstein statistics
it is convenient to decompose the wave function of identical (charged
in our case) bosons with momenta $p_1$ and $p_2$ into the wave
function of the center-of-mass system (c.m.) with total momentum $P =
\frac{1}{2}(p_1 + p_2)$ and the inner wave function with relative
momentum $Q = (p_1 - p_2) = 2q$. It allows us to express $A_{12}$ in
terms of the confluent hypergeometric function $\Phi$~\cite{schiff}:
\begin{eqnarray}
 A_{12} &=& \frac 1{\sqrt{2}} [ \Psi(\vect{q},\vect{r}) +
            \Psi_S(\vect{q},\vect{r}) ]\:, \label{eq:A}\\
 \Psi(\vect{q},\vect{r}) &=& \Gamma(1+i\eta)e^{-\pi \eta/2}
                              e^{i\vecs{q}\cdot\vecs{r}}
           \Phi(-i\eta;1;iqr(1 - \cos \theta))\:,\nonumber\\
 \Psi_S(\vect{q},\vect{r}) &=& \Gamma(1+i\eta)e^{-\pi \eta/2}
                              e^{-i\vecs{q}\cdot\vecs{r}}
                       \Phi(-i\eta;1;iqr(1 + \cos \theta))
                          \:,\nonumber
\end{eqnarray}
where $r = x_1 - x_2$ and the parameter $\eta = m\alpha/2q$. Assuming
factorization in the source functions, $\rho(r_1,r_2)
=\rho(r_1)\rho(r_2) = \rho(R)\rho(r)$ (here $R = \frac{1}{2}(x_1 +
x_2)$), one obtains the following expression for theoretical BEC
formula~\cite{bowler91} including the improved Coulomb correction
\cite{biya95}:
\begin{eqnarray}
N^{(\pm \pm)}/ N^{\rms{BG}} &=& \frac 1{G(q)} \int \rho(R)
                              d^3R \int \rho(r) d^3r |A_{12}|^2 \nonumber\\
 &=& \sum_{n=0}^{\infty} \sum_{m=0}^{\infty} \frac{(-i)^n(i)^m}
     {n+m+1} (2q)^{n+m}I_R(n,m) A_n A_m^* \nonumber\\
 &\times& \left[ 1+ \frac{n!m!}{(n+m)!}
  \left( 1+\frac{n}{i\eta}\right) \left( 1-\frac{m}{i\eta}\right)
  \right ] \label{eq:BEC}\\
  &=& (1 + \Delta_{\rms{1C}}) + (\Delta_{\rms{EC}} +
       E_{\rms{2B}}),\label{eq:result}
\end{eqnarray}
where $G(q) = 2\pi \eta /(e^{2\pi \eta} - 1)$ denotes Gamow factor
and the first and the second parentheses in eq.~(\ref{eq:result})
correspond to the first and the second terms in
eq.~(\ref{eq:BEC})\footnote{For the exact formulae for
$\Delta_{\rms{EC}}$ and $E_{\rms{2B}}$ see \cite{biya95}.}
and
$$
I_R(n,m) = 4 \pi \int dr\, r^{2 + n + m} \rho (r),\qquad
A_n = \frac{\Gamma(i\eta + n)}{\Gamma(i\eta)}\frac{1}{(n!)^2}.
$$

To analyse data corrected only by the Gamow factor using our
formulae we should use the following ratio:
\begin{eqnarray}
  N^{(\pm\pm:\rms{GC})}/N^{\rms{BG}} (Q = 2q) &\!\!\!=&\!\!\!
                                R_{\rms{CC}}/G(q)  \nonumber\\
 &\!\!\!=&\!\!\! c(1 + \Delta_{\rms{1C}} + \Delta_{\rms{EC}})
      \left[1 + \lambda \frac{E_{\rms{2B}}}{1 + \Delta_{\rms{1C}} +
      \Delta_{\rms{EC}}}\right] (1 + \gamma Q).\hspace{1cm}
      \label{eq:ratio}
\end{eqnarray}
It should be noted that the normalization and an effective degree of
coherence, i.e., the denominator of the ratio $E_{\rms{2B}}/( 1 +
\Delta_{\rms{1C}} + \Delta_{\rms{EC}})$, are related to each other.
Notice also that other parameters like the additional normalization
factor $c$, the long range correlation $\gamma$ and $\lambda$ are
introduced by hand.\\

{\bf Source function:}~~~To obtain an explicit expression, we have to
decide on some form of the source function.  In the present Letter,
we shall use the Gaussian source distribution, $\rho(r) =
\frac{\beta^3}{\sqrt{\pi^3}}\exp(-\beta^2 r^2)$.  For this type of
source function we have the following formulae for the elements of
eqs.~(\ref{eq:result}) and (\ref{eq:ratio})~\cite{biya95}:
\begin{eqnarray}
  I_R^{\rms G} (n,m) &=& \frac {2}{\sqrt{\pi}}
                       \left(\frac 1{\beta}\right)^{n+m}
               (n+m+1)~\Gamma\left( \frac{n+m+1}{2} \right),\\
  E_{\rms{2B}} &=& \exp\left(-\frac{q^2}{\beta^2}\right).
\end{eqnarray}

To analyze data corrected by the Coulomb wave function as was done
in~\cite{boggild}, we should modify the formula (\ref{eq:ratio})
replacing it by the following one:
\begin{eqnarray}
  N^{(\pm\pm:\rms{CC})}/N^{\rms{BG}} (Q = 2q) &\!\!\!=&\!\!\!
                                           \frac{R_{\rms{CC}}}
     {G(q)(1 + \Delta_{\rms{1C}} + \Delta_{\rms{EC}})}\nonumber\\
  &\!\!\!=&\!\!\! c\left[1 + \lambda \frac{E_{\rms{2B}}}{1 +
                   \Delta_{\rms{1C}} + \Delta_{\rms{EC}}} \right]
                   (1 + \gamma Q).  \label{eq:a21}
\end{eqnarray}

For the sake of reference, we write down here also the conventional
formula (i.e., a kind of standard formula without corrections due
to the final state interactions):
\begin{equation}
  N^{(\pm\pm:\rms{Standard})}/N^{\rms{BG}} (Q = 2q)
         = c\left[1 + \lambda E_{\rms{2B}} \right]
           (1 + \gamma Q).         \label{eq:a22}
\end{equation}

{\bf Asymptotic expansion of the Coulomb wave function:}~~~First of
all, it should be noted that the expansion in eq.~(\ref{eq:result})
has to be the limited to $q_{\rms{limit}} ( = Q/2)$ only due to
mathematical properties of the confluent hypergeometric function
used~\cite{schiff}. This can be seen as wild oscillation developing
in Fig.~1~(a) where the eq.~(\ref{eq:ratio}) has been simply used. If
we, instead, set the Coulomb correction to zero in the region $q >
q_{\rms{limit}}$ limit, a small step appears as seen in Fig.~1~(b).
Therefore in order to analyse in a consistent way the whole region of
the momentum transfer measured we have to use the following
asymptotic expansion of the Coulomb wave function:
\begin{eqnarray}
 A_{12}^{\rms{asym}} &\!\!\! = &\!\!\! \frac 1{\sqrt{2}} \left[
                       \Psi^{\rms{asym}} (\vect{q},\vect{r}) +
                       \Psi_{\rms S}^{\rms{asym}}(\vect{q},\vect{r})
                        \right] \:, \label{eq:asym}\\
  \Psi^{\rms{asym}} (\vect{q}, \vect{r}) &\!\!\! = &\!\!\!
             \exp\{i(qz - \eta\ln(r-z))\}\times \left(1+\frac{\eta^2}
              {iq(r-z)}\right)\nonumber\\
  &\!\!\! &\!\!\! + f(\theta) \frac{\exp\{i(qr - \eta \ln(2qr))\}}r\:,
                                \label{eq:asym1}\\
  \Psi_{\rms S}^{\rms{asym}} (\vect{q}, \vect{r}) &\!\!\! = &\!\!\!
                      \exp\{i(-qz - \eta
               \ln(r+z))\}\times \left(1+\frac{\eta^2}{iq(r+z)}\right)
                      \nonumber\\
  &\!\!\! &\!\!\! + f(\pi - \theta)
              \frac{\exp\{i(qr - \eta \ln(2qr))\}}r\:, \label{eq:asym2}
\end{eqnarray}
where $z = r \cos \theta$ and
$$
f(\theta) = - \frac{\eta}{2q}\frac 1{\sin^2 (\theta/2)}
\exp\{-2i\eta\ln \sin (\theta/2) + 2i \arg \Gamma (1+i\eta)\} .
$$
In analyses we should assure a smooth connection between both regions
of $q = Q/2$. To avoid the divergence of denominators $(1 \pm
\cos\theta)$, we have to introduce a cutoff parameter $\epsilon$ (of
the order of $ \epsilon \simeq 10^{-3}$) such that $(1 \pm \cos\theta)
> \epsilon$ always\footnote{We have confirmed that the parameter $\epsilon$
depends on the magnitude of the interaction region ($R$).}.  This
procedure, shown as a flow chart in fig.~2, is called the ``seamless
fitting~(SF)''.\\

{\bf Analyses of data by SF method:}~~~Our results obtained in terms
of the new formula (\ref{eq:ratio}) are shown in Table~I and Figs.~3
and 4. Whereas the parameter $\gamma$ (representing here the
influence of long range correlations) increases now noticeably in
comparison with that obtained previously in
\cite{biya94d}\footnote{This shows that Coulomb corrections can be
important in the bigger range of momentum transfers than considered
so far and justifies {\it a posteriori} our present investigation.},
we find that the interaction region represented by $R = 1/2\beta$
remains S + Pb reactions almost the same. Only in $p$ + Pb collisions
the estimated $R$ becomes larger with inclusion of Coulomb correction
than that obtained in \cite{biya94d}. These facts suggest that we
should be careful in interpreting any data (at least those for kaons)
which were corrected for Coulomb interactions only by the Gamow
factor.\\

{\bf Concluding remarks:}~~~We have proposed the possible method of
applying the Coulomb correction for the BEC in the whole region of
momentum transfer, the ``seamless fitting" (SF).  We confirm that
this method works well when analysing data corrected only by the
Gamow factor. \\

Our analyses of data of $KK$ pairs in S + Pb reaction \cite{na4494}
shows (cf. Table~I) that $R(K^+K^+)$ = 4 fm and $R(K^-K^-)$ = 3 fm
(i.e., they differ substantially), contrary to the estimation
provided in \cite{na4494} that $R(K^{\pm}K^{\pm})$ $\approx$ 3 fm.
This is an important result for the study of signals of the Quark
Gluon Plasma (QGP) (see refs. \cite{biya94,biya94d,nagamiya92}).
Moreover, we found that the long range correlations are strongly
affected by Coulomb corrections (most probably because of the long
range character of Coulomb interactions).\\

For the sake of reference, we show also in Table~II and Fig. 5
results of our analysis of data for S + Pb $\to \pi\pi + X $ reaction
\cite{boggild} (which were corrected by the Coulomb wave function
method \cite{pcz90}) performed by using both eq.~(\ref{eq:a21}) and
the  ``standard'' formula (eq.~(\ref{eq:a22})).  As one can see there
are no significant differences between parameters estimated by means
of these two formulae, in particular the magnitude of the interaction
region is in both cases almost the same.\\

{\bf Acknowledgements:}~~~The authors would like to thank S. Esumi, T.
Nishimura, and S. D. Pandey (members of NA44 Collaboration) for useful
conversations and correspondences.  This work is partially supported by
Japanese Grant-in-Aid for Scientific Research from the Ministry of Education,
Science and Culture (\# 06640383).
%
%   REFERENCES
%
%\newpage

%
%   TABLE I
%
%\newpage
\begin{table}[htbp]
\caption[table1]{Parameters of kaonic BEC obtained from S + Pb and $p$
 + Pb reactions.}
\label{tbl:a1}
\begin{center}
\begin{tabular}{c|c|ccccc}
\hline \hline
reaction
& formula
& c
& $R$ [fm]
& $\lambda$
& $\gamma$
& $\chi^2/$NDF\\
\hline
& \lw{standard}
&  0.998$\pm$0.008
&  3.384$\pm$0.171
&  1.010$\pm$0.071
&  ---
&  58.0/33\\
S + Pb $\to$
&
&  1.101$\pm$0.022
&  3.991$\pm$0.234
&  0.977$\pm$0.078
& $-$0.420$\pm$0.079
&  36.8/32\\
\cline{2-7}
$K^+K^+ + X$
& \lw{eq.~(4)}
&  0.966$\pm$0.007
&  3.918$\pm$0.000
&  0.804$\pm$0.059
&  ---
&  30.9/33\\
&
&  0.987$\pm$0.015
&  4.285$\pm$0.000
&  0.825$\pm$0.074
& $-$0.071$\pm$0.077
&  30.9/32\\
\hline
& \lw{standard}
&  0.988$\pm$0.021
&  3.058$\pm$0.285
&  1.070$\pm$0.142
&  ---
&  31.9/32\\
S + Pb $\to$
&
&  0.989$\pm$0.066
&  3.061$\pm$0.345
&  1.069$\pm$0.159
& $-$0.004$\pm$0.279
&  31.9/31\\
\cline{2-7}
$K^-K^- + X$
& \lw{eq.~(4)}
&  0.944$\pm$0.017
&  3.089$\pm$0.000
&  0.836$\pm$0.111
&  ---
&  35.0/32\\
&
&  0.872$\pm$0.048
&  3.066$\pm$0.012
&  1.030$\pm$0.178
&  0.411$\pm$0.277
&  32.5/31\\
\hline
& \lw{standard}
&  0.989$\pm$0.027
&  2.134$\pm$0.291
&  0.636$\pm$0.089
&  ---
&  34.3/26\\
$p$ + Pb $\to$
&
&  1.149$\pm$0.078
&  2.698$\pm$0.559
&  0.493$\pm$0.114
& $-$0.546$\pm$0.237
&  30.9/25\\
\cline{2-7}
$K^+K^+ + X$
& \lw{eq.~(4)}
&  0.975$\pm$0.019
&  2.731$\pm$0.000
&  0.520$\pm$0.091
&  ---
&  32.7/26\\
&
&  0.996$\pm$0.061
&  2.310$\pm$0.001
&  0.431$\pm$0.133
& $-$0.180$\pm$0.256
&  30.7/25\\
\hline
\end{tabular}
\end{center}
\end{table}
%
%   TABLE II
%
\begin{table}[htbp]
\caption[table2]{Parameters of S + Pb $\to \pi^{\pm}\pi^{\pm} + X$ reactions.}
\label{tbl:a2}
\begin{center}
\begin{tabular}{c|ccccc}
\hline \hline
formula
& c
& $R$ [fm]
& $\lambda$
& $\gamma$
& $\chi^2/$NDF\\
\hline
\lw{standard}
&  0.800$\pm$0.003
&  4.506$\pm$0.327
&  0.461$\pm$0.042
&  ---
&  17.5/16\\
&  0.824$\pm$0.010
&  5.015$\pm$0.421
&  0.450$\pm$0.044
& $-$0.178$\pm$0.065
&  10.9/15\\
\hline
\lw{eq.~(7)}
&  0.800$\pm$0.003
&  4.575$\pm$0.335
&  0.509$\pm$0.048
&  ---
&  17.8/16\\
&  0.825$\pm$0.010
&  5.105$\pm$0.436
&  0.502$\pm$0.052
& $-$0.180$\pm$0.065
&  11.0/15\\
\hline
\end{tabular}
\end{center}
\end{table}
%
%   FIGURE CAPTIONS
%
%\newpage
\section*{{\large\bf Figure Captions}}
\begin{description}
  \item[Fig.~1. ] (a) Analyses of data for S + Pb $\to K^-K^- + X$
                  reaction by eq.~(\ref{eq:ratio}). (b) The same but with
                  Coulomb corrections switched off for
                  $q > q_{\rms{limit}} = Q/2$ limit. (Results for
                  $K^+K^+$ pair production looks similar with only
                  change being in the value of $q_{\rms{limit}}$
                  which depend on the value of radius parameter $R$.)
  \item[Fig.~2. ] Flow chart for our procedure of ``seamless fitting~(SF)''.
  \item[Fig.~3. ] Results of SF for BEC for kaons produced in S + Pb
                  collisions; (a) for $K^+K^+$ pairs; (b) for $K^-K^-$
                  pairs.
  \item[Fig.~4. ] Results of SF for BEC for kaons produced in
                  $p$ + Pb $\to K^-K^- + X$ reaction.
  \item[Fig.~5. ] Results of SF for BEC for pions produced in
                  S + Pb $\to \pi^+\pi^+ + X$ reaction.
\end{description}
\end{document}